\documentclass[nohyper]{JHEP3}

\newcommand{\AAA}{\mbox{\boldmath$A$}}
\newcommand{\BBB}{\mbox{\boldmath$B$}}
\newcommand{\CCC}{\mbox{\boldmath$C$}}
\newcommand{\DDD}{\mbox{\boldmath$D$}}

\title{Real and Imaginary Mass Generation in the Presence of
External Fields and Axions}

\author{Stefano Ansoldi\\
Dipartimento di Fisica Teorica, Universit\`a di Trieste
and INFN, Sezione di Trieste, Strada Costiera, 11 - I-34014 Miramare, Trieste, Italy\\
\email{ansoldi@trieste.infn.it}}

\author{Eduardo I. Guendelman\\
Physics Department, Ben Gurion University, Beer Sheva, Israel\\
\email{guendel@bgumail.bgu.ac.il}}

\author{Euro Spallucci\\
Dipartimento di Fisica Teorica, Universit\`a di Trieste
and INFN, Sezione di Trieste, Strada Costiera, 11 - I-34014 Miramare, Trieste, Italy\\
\email{spallucci@trieste.infn.it}}

\abstract{Small fluctuations around a constant electric or constant
magnetic field $F$ are analyzed in a theory with pseudo scalar
$\phi$ with a coupling $g \phi F \tilde{F}$. It is found
that a magnetic external field leads to mass generation
for the small perturbations, while an electric field
suffers from a tachyonic mass generation in the case in which
the field strength is higher than a critical value
(related to the pseudo scalar mass). The vacuum energy can be
exactly evaluated and it is found that an imaginary part is
present when the external electric field exceeds its critical
value.}

\keywords{Mass generation, Non-perturbative effects, Axions, Pseudoscalar coupling}

\preprint{}

\begin{document}

\section{Introduction}

\subsection{Preliminary considerations}

A dynamical mechanism providing mass to vector gauge bosons is instrumental
to match theoretical models of fundamental interactions with the particle
spectrum observed in high energy experiments. A blueprint of dynamical
mass generation is given by the Schwinger Model, or QED$_{2}$, where fermions
quantum fluctuations induce a mass term for the two-dimensional photon.
Extension of this non-perturbative quantum effect to four-dimensional gauge
theories has still to come because the gauge field effective action cannot be
computed exactly in $4D$. In the meanwhile, the archetypal mechanism for
gauge field theory  mass generation is Spontaneous Symmetry Breaking,
induced either by classical tachyonic mass terms \cite{bib:higgs} or by
quantum
radiative corrections \cite{bib:cw}. The Coleman-Weinberg breaking of gauge
symmetry avoids classical tachyonic mass terms and gives raise to a
non-vanishing
vacuum expectation value for massless scalar fields through radiative quantum
corrections.
In this paper we are going to discuss a ``complementary'' mechanism, where
mass follows from the breaking of rotational invariance induced by a classical
background configuration of the gauge field strength. A real, or tachyonic,
mass is obtained according with the magnetic, or electric, nature of the
background field. The model implementing this effect consists of
a scalar field $\phi$ non-minimally coupled to a $U(1)$ gauge vector
(non-abelian extensions of this model are planned for future investigations)
through an interaction term of the form:
\begin{equation}
    {\mathcal{L}} _{\mathrm{I}}
    =
    \frac{g}{8}\phi
    \epsilon _{\mu \nu \alpha \beta} F ^{\mu \nu} F ^{\alpha \beta}
    .
\label{eq:psescacou}
\end{equation}
The interaction term $ {\mathcal{L}} _{\mathrm{I}}$ has a long history
dating back to the celebrated ABJ anomaly and neutral pion electromagnetic
decay \cite{bib:abj}. Moreover by keeping fixed the form and varying the strength of the
coupling constant, $ {\mathcal{L}} _{\mathrm{I}}$ is equally well suited
to describe  the axion field currently appearing in many astrophysical and
quantum field theoretical problems \cite{bib:axiref}.

In what follows we are going to analyze the case in which the electromagnetic
field is a purely electric or purely magnetic background, with special interest
about the dynamics of its fluctuations. Before embarking this program and before
giving a more detailed account of the main aspects of our approach, it is worth
to recall some important steps already taken in the past in this direction. In the
main part of this work we will stress more carefully analogies, as well as differences,
with what we are proposing in this paper.
In particular the fact that an external \textit{magnetic}
field modifies the dispersion relation of photons coupled to (pseudo)scalars
was already discussed, for example, in \cite{bib:MaiPetZav}; there the authors
have in mind an experimental set-up for the detection of pseudoscalars coupled
to two photons based on the fact that the photon effective mass provided by the
pseudoscalar coupling is responsible for an ellipticity in an initially
linearly polarized beam\footnote{This set-up is now a fully working project,
PVLAS, at the I.N.F.N. Legnaro laboratories, taking data since May 1999: about this
we refer the interested reader to \cite{bib:QED2000} and references therein, as well
as to \cite{bib:PVLASAxi} for a more direct connection with axion experiments.}.
Concerning the situation in which a background electric field is present, recently
this problem has been analyzed in \cite{bib:GriMasMoh}, where the authors show under
which conditions an external electric field decays to pseudoscalars and discuss
some particular configurations in which their results can be applied.
Postponing a deeper analysis to what follows, we think that an important point to be
stressed already at this early stage, is the fact that in the above studies the
discussion is \textit{perturbative} whereas, in the present paper, we are going to analyze a
second order effective approximation for the dynamics of the fluctuations of the
electromagnetic field, only after a
full, \textit{non-perturbative} treatment of the pseudoscalar.

Before developing this part, we will shortly present some interesting features
of the model in a \textit{naive} form. We remember that, indeed,
it is a special feature of the coupling (\ref{eq:psescacou})
to generate physical masses, or tachyonic instabilities,
when an appropriate classical configuration of the scalar field $\phi$ is
turned on: performing an integration by parts in the action
associated with (\ref{eq:psescacou}), we end up with an interaction term
of the form
\begin{equation}
    {\mathcal{L}} _{\mathrm{I}}
    =
    \frac{g}{8}
    \left( \partial _{\beta} \phi \right)
    \epsilon ^{\mu \nu \alpha \beta} F _{\mu \nu} A _{\alpha}
    .
\label{eq:psescacoumas}
\end{equation}
We can then consider a background configuration selecting a preferred spacelike
direction, e.g.
$\langle \phi \rangle = \mathrm{const.} \cdot \delta^3_\mu x ^\mu$, so that
(\ref{eq:psescacoumas}) gives a $(2+1)$-dimensional, Chern--Simons type,
mass term \cite{bib:CheSim}. The resulting massive Chern--Simons model is
embedded into a $(3+1)$-dimensional theory. Thus, rotational invariance in
embedding space is broken \cite{bib:MasGenRotInvBre}. It is interesting to
compare this result with the case in which the background field is time dependent
only, i.e.  $\langle \phi \rangle \equiv \varphi (t)$. The resulting
Chern--Simons model is endowed with a tachyonic mass term \cite{bib:tacmasgen},
meaning that this type of background field is unstable.
This kind of instabilities could
play a role in baryogenesis, as it has been argued in
\cite{bib:tacmasgen}, \cite{bib:bargen}. Moreover,
a time dependent axion field may also affect the growth of primordial
magnetic fields \cite{bib:primagfie}, produce effective Lorentz and parity
violating modifications of electrodynamics and affect the polarization
of radiation coming from distant galaxies \cite{bib:poldisgal}.

\subsection{A ``naive'' demonstrative approach}

In this paper we want to explore ``the other side of the Moon'',
that is what happens if the gauge field strength $F _{\mu \nu}$, and not
the scalar/axion field $\phi$, acquires a non-vanishing background value.
As an example, suppose
$\langle F _{\mu \nu} \rangle = B \delta _{[ \mu \vert 2} \delta _{\nu ] 3}$,
i.e.  $\langle F _{\mu \nu} \rangle$ is a constant magnetic field along the
$x ^{3}$ direction. This choice is not \textit{ad hoc} as it could appear: we are
trying to mimic  the QCD vacuum, where
a constant, color, magnetic field lowers the energy density with respect to
the perturbative Fock vacuum, where no gluons are present \cite{bib:qcd}.
By introducing the proper kinetic term for $\phi$ and the
fluctuation field $f_{\mu\nu}$, we see  that (\ref{eq:psescacoumas})
turns into
\begin{equation}
    {\mathcal{L}}
    =
    - \frac{1}{4} f _{\mu \nu} f ^{\mu \nu}
    + \frac{1}{2} \partial _{\mu} \phi \partial ^{\mu} \phi
    + \frac{g}{4} B\phi \epsilon _{2 3 \alpha \beta} f ^{\alpha \beta}
    .
\label{eq:psescamaglag}
\end{equation}
Equation (\ref{eq:psescamaglag}) shows as $\phi$ couples only to the $f ^{01}$
component of the fluctuation field strength. Accordingly, we can write the
following effective Lagrangian in the $(0-1)$-plane:
\begin{equation}
    {\mathcal{L}}_{01}
    \equiv
    - \frac{1}{2} f _{01} f ^{01}
    + \frac{1}{2} \partial _{\mu} \phi \partial ^{\mu} \phi
    + \frac{g }{4} B \phi f ^{01}
    .
\label{eq:l01}
\end{equation}
By freezing the $x^2$, $x^3$ dependence of the fields in (\ref{eq:l01}) we
get  the bosonized Schwinger model \cite{bib:BosSchMod} and solving
the classical field equation for $\phi$ we obtain the non-local effective
action
for the dimensionally reduced, massive, electromagnetic field
\begin{equation}
    {\mathcal{L}}_{01}
    \equiv
    - \frac{1}{2} f _{01} f ^{01}
    - \frac{g^2 B^2}{32} f ^{01}\frac{1}{\Box}f _{01}
    ,
\label{eq:l01emf}
\end{equation}
where the generated mass square is proportional to $g^2 B^2$.
Of course the full spectrum of excitations will contain more
general types of spacetime dependence, so that we expect
here an anisotropic mass generation effect.\\
We can repeat the same calculations with an electric background, e.g.
$$
    \langle F _{\mu \nu} \rangle = E \delta _{[ \mu \vert 0}
    \delta _{1 \vert \nu ]}
    .
$$
Then, instead of (\ref{eq:l01emf}), we obtain for the small
excitations in the  $(2-3)$-plane
\begin{equation}
    {\mathcal{L}}
    =
    - \frac{1}{2} f _{23} f ^{23}
    + \frac{g^2 E^2}{32} f ^{23} \frac{1}{\Box} f _{23}
    .
    \label{eq:l02}
\end{equation}
By comparing (\ref{eq:l02}) with (\ref{eq:l01emf}),  we see that in the electric
case there is a different sign in front of the non-local term, i.e.
the generated mass is tachyonic. This situation for the electric background
can be partially cured providing a non vanishing mass $m _{\mathrm{A}}$
to the axion field, so that the tachyonic mass generation sets in
only around the critical electric field  $g E \sim m _{\mathrm{A}}$.

As we will see
in the next sections, the theory can be solved beyond the two
dimensional truncated sectors mentioned here. Furthermore, it is possible to
compute the vacuum energy  resulting from small fluctuations around the
chosen background. The vacuum energy develops an imaginary part
for electric fields bigger than some critical value, of the order of
$m _{\mathrm{A}} / g$. It can be worth now
to remark a difference with respect to the instability in two dimensional
QED. Both effects are triggered by  an
external electric field, but the standard Schwinger mechanism
operates for any value of the constant electric field, provided it extends
to sufficiently large distances. Even with a weak field, one can
perform enough work to create a new pair. By contrast, in
our case, the effect is present only for field strengths bigger than a
critical one, regardless of whether it extends over a large region of
space or not.

The paper is organized as follows.
In Subsection \ref{sec:patint}
we set up the ``sum over histories'' formulation for the system under study:
in particular we show how an effective Lagrangian for small
fluctuations around a pure electric/magnetic background can be obtained,
as a quadratic approximation, after integrating out exactly (and \textit{non-perturbatively})
the scalar field
and neglecting higher order terms. The next Subsection (\ref{sec:sigmasgen})
deals in more detail with the mass generation effect, fully
exploiting the difference
between the electric and the magnetic case and determining the propagator and its
spectrum in momentum space. Then in Section \ref{sec:proandvacene} the expression
of the vacuum energy is introduced together with
the exact expression for the propagator. The main steps in the computation
of the free energy are outlined in Section \ref{sec:vaceneeva} and the
emergence of an imaginary part in the pure electric case is analyzed.
Discussion and conclusions are in Section \ref{sec:discon}, with particular emphasis
about the relation between the tachyonic mass-generation effect and a possible instability
of homogeneous electric fields above some threshold. Three appendices
follow, where technical details can be found about the solution of the
eigenvalue equation for the momentum space propagator (Appendix \ref{app:appdiaeig}),
the determination of the propagator itself (Appendix \ref{app:protecdet}) and the
explicit computation of the free energy (Appendix \ref{app:vacenecomtecdet}), with
particular attention at the behavior in the infrared and ultraviolet limits.

\section{\label{sec:genana}General analysis of the small perturbations}

\subsection{\label{sec:patint}Path integral}

An effective technical framework to investigate quantum
fluctuations around an external background configuration is provided
by the Feynman path integral formalism.
The partition function(al) encoding the
dynamics of the interacting $\phi$ and $A _{\mu}$ fields reads
\begin{equation}
    Z
    \equiv
    \int [ {\mathcal{D}} \phi ] [ {\mathcal{D}} A ]
    \exp\left\{ - \imath \int d ^{4} x {\mathcal{L}} \right)
    \label{pathint}
    ,
\end{equation}
where
\begin{equation}
    {\mathcal{L}}
    =
    -
    \frac{1}{4}
    F _{\mu \nu}\, F ^{\mu \nu}
    +
    \frac{g}{8}\,\phi\,
    \epsilon ^{\mu \nu \rho \sigma} F _{\mu \nu}\, F _{\rho \sigma}
    +
    \frac{1}{2}
    \partial _{\mu} \phi\, \partial ^{\mu} \phi
    + \frac{m ^{2} _{\mathrm{A}}}{2} \phi ^{2}
    \label{eq:StartingL}
\end{equation}
and both the gauge fixing and ghost terms are, momentarily, understood in the
functional measure $[{\mathcal{D}} A]$.
 For the sake of generality, we assigned a non-vanishing mass to the pseudo
scalar field $\phi$.
Since the path integral (\ref{pathint}) is gaussian in $\phi$, the scalar
field can be integrated away exactly, i.e.
\begin{eqnarray}
    & &
    \int  [ {\mathcal{D}} \phi ]
        \exp
        \left\{
            - \imath
            \int d ^{4} x
                \left[
                    \frac{1}{2}
                    \partial _{\mu} \phi
                    \partial ^{\mu} \phi
                    +
                    \frac{g}{8}
                    \phi
                    \varepsilon ^{\mu \nu \rho \sigma}
                    F _{\mu \nu} F _{\rho \sigma}
                    +
                    \frac{m _{\mathrm{A}} ^{2}}{2} \phi ^{2}
                \right]
        \right\}
    =
    \nonumber \\
    & & \quad
    =
    \left[
        \det
            \left(
                \frac{ \Box + m ^{2} _{\mathrm{A}}}
                     {\mu ^{2}}
            \right)
    \right] ^{-1/2}
    \! \! \! \! \!
    \exp
    \left\{
        \imath
        \frac{g^2}{128}
        \int d ^{4} x
            \epsilon ^{\mu \nu \rho \sigma}
            F _{\mu \nu} F_{\rho \sigma}
            \frac{1}{\Box + m ^{2}_{\mathrm{A}}}
            \epsilon ^{\alpha \beta \gamma \delta}
            F _{\alpha\beta}
            F_{\gamma\delta}
    \right\}
    ,
    \label{eq:NonPer}
\end{eqnarray}
where $\mu$ is a mass scale coming from the definition of the measure
$[ {\mathcal{D}} \phi ]$.
Integrating out the $\phi$ field induces a
non-local effective action for the $A$ field.
We stress that this effective action, being obtained at the \textit{non-perturbative}
level, takes into account (as an effective action for $A$) all the effects due to the
presence of the pseudoscalars at all perturbative orders. This is a crucial difference
with many of the previous works on the subject. As is clear from (\ref{eq:NonPer}),
the resulting path integral
is quartic in $A$ but, even if it cannot be computed in a closed form,
the background field method provides a reliable approximation scheme to
deal with this problem.
We thus split $F _{\mu \nu}$ in the sum of a classical background
$\langle F _{\mu \nu} \rangle$ and a small fluctuation $f _{\mu \nu}$:
\begin{equation}
    F _{\mu \nu}
    =
    \langle F _{\mu \nu} \rangle
    +
    f _{\mu \nu}
    .
\end{equation}
In the case of a pure electric or a pure magnetic background we have
\begin{equation}
    \epsilon ^{\mu \nu \alpha \beta}
    \langle F _{\mu \nu} \rangle
    \langle F _{\alpha \beta} \rangle
    =
    0
    .
\end{equation}
Thus, expanding ${\mathcal{L}}$ up to quadratic terms and dropping
a total divergence, we obtain
for the effective Lagrangian of the fluctuations of the electromagnetic field
\begin{equation}
    {\mathcal{L}} ^{(2)}
    =
    -
    \frac{1}{4}
    \langle F _{\mu \nu} \rangle
    \langle F ^{\mu \nu} \rangle
    -
    \frac{1}{4}
    f _{\mu \nu}
    f ^{\mu \nu}
    -
    \frac{g ^{2}}{16}
    \epsilon ^{\mu \nu \alpha \beta}
    \langle F _{\mu \nu} \rangle
    \epsilon ^{\rho \sigma \gamma \delta}
    \langle F _{\rho \sigma} \rangle
    f _{\alpha \beta}
    \frac{1}{\Box + m ^{2} _{\mathrm{A}}} f _{\gamma\delta}
    \label{eq:LquadraticElectric}
    .
\end{equation}
As we will see in a while, there is an important sign difference between the
magnetic and the electric case.

\subsection{\label{sec:sigmasgen}The signature of the mass generation:
ordinary \textit{versus} tachyonic}

In this subsection we turn to the analysis of the contribution
\begin{equation}
    \epsilon ^{\mu \nu \alpha \beta}
    \langle F _{\mu \nu} \rangle
    \epsilon ^{\rho \sigma \gamma \delta}
    \langle F _{\rho \sigma} \rangle
\end{equation}
of equation (\ref{eq:LquadraticElectric}), which will be responsible
for ordinary or tachyonic mass generation according to whether
$\langle F _{\mu \nu} \rangle$ represents an external magnetic
or electric field. We notice that our results are not inconsistent with those that
can be deduced from equation (31) of \cite{bib:GriMasMoh}, from which it is clear
that the sign of the contribution from the interaction Lagrangian changes in the case
of purely electric or purely magnetic background.

Let us now start from the  case of a constant magnetic field $B$. Then, without
loosing generality, we can rotate the reference frame to align an axis, say
$x ^{1}$, with $B$. Accordingly,
$\langle F _{\mu \nu} \rangle = B \delta _{[ \mu | 2} \delta _{ \nu ] 3}$,
so that
\begin{equation}
    \epsilon ^{\mu \nu \alpha \beta}
    \langle F _{\mu \nu} \rangle
    \epsilon ^{\rho \sigma \gamma \delta}
    \langle F _{\rho \sigma} \rangle
    =
    4 B ^{2}
    \epsilon ^{2 3 \alpha \beta}
    \epsilon ^{2 3 \gamma \delta}
    \label{eq:Fmnmag}
    .
\end{equation}
It is clear thus that $\alpha$, $\beta$, $\gamma$ and $\delta$
in (\ref{eq:Fmnmag}) can take only the values $0$, $1$. Denoting
by ${}^{(2)} \eta ^{\alpha \beta}$ the $2 \times 2$ Minkowski tensor,
we must then have that
\begin{eqnarray}
    \epsilon ^{\alpha \beta 2 3}
    \epsilon ^{\gamma \delta 2 3}
    & = &
    A
    \left(
        {}^{(2)} \eta ^{\alpha \gamma} {}^{(2)} \eta ^{\beta \delta}
        -
        {}^{(2)} \eta ^{\alpha \delta} {}^{(2)} \eta ^{\beta \gamma}
    \right)
    \nonumber \\
    & = &
    A
    \left(
        P _{(10)} ^{\alpha \gamma} P _{(10)} ^{\beta \delta}
        -
        P _{(10)} ^{\alpha \delta} P _{(10)} ^{\beta \gamma}
    \right)
    \label{eq:epspromagcas}
    ,
\end{eqnarray}
where $P _{(10)} ^{\alpha \gamma}$ is the projector onto the $(0-1)$-plane,
i.e. $P _{(10)} ^{\alpha \gamma} = {}^{(2)} \eta ^{\alpha \gamma}$.
Contracting in relation (\ref{eq:epspromagcas}) the couples
of indices $({}^{\alpha \gamma})$ and $({}^{\beta \delta})$ we obtain
$$
    \epsilon ^{\alpha \beta 2 3}
    \epsilon _{\alpha \beta} {}^{2 3}
    =
    A (2 \times 2 - 2)
    =
    2 A
    ;
$$
since in lowering the indices $\alpha \beta$ there is the time
involved,
$
    \epsilon ^{\alpha \beta 2 3}
    \epsilon _{\alpha \beta} {}^{2 3}
    =
    - 2
$,
so that $A = -1$.

Analogously in the electric case we can take
$F _{\mu \nu} = E \delta _{[\mu | 0} \delta _{1 | \nu]}$. Then
\begin{equation}
    \epsilon ^{\alpha \beta \mu \nu}
    \langle F _{\mu \nu} \rangle
    \epsilon ^{\gamma \delta \rho \sigma}
    \langle F _{\rho \sigma} \rangle
    =
    4 E ^{2}
    \epsilon ^{\alpha \beta 0 1}
    \epsilon ^{\gamma \delta 0 1}
    \label{eq:Fmnele}
    ,
\end{equation}
with the indices $\alpha$, $\beta$, $\gamma$, $\delta$ taking only the
\textit{spatial} values $2$, $3$. Now, following the same procedure and observing
that no minus signs are involved in lowering spatial indices,
we find
\begin{eqnarray}
    \epsilon ^{\alpha \beta 0 1}
    \epsilon ^{\gamma \delta 0 1}
    & = &
    \left(
        {}^{(2)} \delta ^{\alpha \gamma} {}^{(2)} \delta ^{\beta \delta}
        -
        {}^{(2)} \delta ^{\alpha \delta} {}^{(2)} \delta ^{\beta \gamma}
    \right)
    \nonumber \\
    & = &
    P ^{\alpha \gamma} _{(23)} P ^{\beta \delta} _{(23)}
    -
    P ^{\alpha \delta} _{(23)} P ^{\beta \gamma} _{(23)}
    \nonumber
    ;
\end{eqnarray}
again we introduced the projector notation
$P ^{\mu \nu} _{(23)} \equiv {}^{(2)} \delta ^{\mu \nu}$, where
${}^{(2)} \delta ^{\mu \nu}$ is the $2 \times 2$ Kronecker delta.

We can now write the equation of motion for the fluctuations, which
from the action (\ref{eq:LquadraticElectric}) turns out to be
\begin{equation}
    \partial _{\mu}
    f ^{\mu \beta}
    =
    8
    g ^{2}
    \epsilon ^{\mu \nu \alpha \beta}
    \langle F _{\mu \nu} \rangle
    \frac{1}{\Box + m _{\mathrm{A}} ^{2}}
    \epsilon ^{\rho \sigma \gamma \delta}
    \langle F _{\rho \sigma} \rangle
    \partial _{\alpha} f _{\gamma \delta}
    \label{eq:fieequonesub}
    .
\end{equation}
The above equation can then be expressed, for both
the electric and magnetic case, in Fourier space and in terms of the
vector potential of the fluctuations, which we will call $a _{\mu}$. Since
$f _{\mu \nu} = \partial _{\mu} a _{\nu} - \partial _{\nu} a _{\mu}$,
we have
\begin{equation}
    k _{\mu}
    \left(
        k ^{\mu} a ^{\nu} - k ^{\nu} a ^{\mu}
    \right)
    =
    \frac{\kappa}{k ^{2} - m _{\mathrm{A}} ^{2}}
    \left(
        \bar{k} ^{2} \bar{g} ^{\mu \nu} - \bar{k} ^{\mu} \bar{k} ^{\nu}
    \right)
    a _{\mu}
    .
    \label{eq:fieequonesubFouspa}
\end{equation}
Here and in what follows we define
\begin{eqnarray}
    \bar{g} ^{\mu \nu} & = & P ^{\mu \nu} _{( \cdot \cdot )}
    \label{eq:norprodefuno}
    \\
    \bar{k} ^{\mu}
    & = &
    P ^{\mu \alpha} _{( \cdot \cdot )} k _{\alpha}
    =
    \bar{g} ^{\mu \alpha} k _{\alpha}
    \label{eq:norprodefdue}
\end{eqnarray}
and
$$
    P ^{\mu \alpha} _{( \cdot \cdot )} = P ^{\mu \alpha} _{(10)}
    , \quad
    \kappa = + 32 g ^{2} B ^{2}
$$
in the magnetic case, or
$$
    P ^{\mu \alpha} _{( \cdot \cdot )} = P ^{\mu \alpha} _{(23)}
    , \quad
    \kappa = - 32 g ^{2} E ^{2}
$$
in the electric case.
Equation (\ref{eq:fieequonesubFouspa}) can be written also as
$$
    {{\mathcal{D}} ^{-1}} ^{\mu \nu} ( k )
    a _{\mu}
    =
    0
    ,
$$
where we have defined
\begin{equation}
    {{\mathcal{D}} ^{-1}} ^{\mu \nu} ( k )
    =
    \left(
        k ^{2} g ^{\mu \nu} - k ^{\mu} k ^{\nu}
    \right)
    -
    \frac{\kappa}{k ^{2} - m _{\mathrm{A}} ^{2}}
    \left(
        \bar{k} ^{2} \bar{g} ^{\mu \nu} - \bar{k} ^{\mu} \bar{k} ^{\nu}
    \right)
    .
\label{eq:opedef}
\end{equation}
The operator in (\ref{eq:opedef}) can be diagonalized\footnote{We
will summarize here the final result about eigenvectors and
the corresponding eigenvalues with all the relevant definitions,
referring the reader to the mentioned appendix for the
detailed computation.} as shown in appendix \ref{app:appdiaeig}.
Four linearly independent
physical states satisfy the eigenvector equation of
${{\mathcal{D}} ^{-1}} ^{\mu \nu} (k)$; we will
call them $k _{\mu}$, $\tilde{k} _{\mu}$, $\tilde{k} ^{\perp} _{\mu}$, $E
^{\mu}$.
Their definitions and corresponding eigenvalues are enlisted in table \ref{tab:eigen}.
\TABLE{%
\begin{tabular}{r|l}
        \textbf{Eigenvalues}
        &
        \textbf{Eigenvectors}
        \\ \hline \hline
        \vbox{
        \hbox{\vspace{1mm}}
        \hbox{$0 \phantom{_{\mu}}$}
        }
        &
        \vbox{
        \hbox{\vspace{1mm}}
        \hbox{$k _{\mu}$}
        }
        \\[1mm] \hline
        \vbox{
        \hbox{\vspace{1mm}}
        \hbox{
        ${\displaystyle{}k ^{2} - \frac{\kappa \bar{k} ^{2}}{k ^{2} - m _{\mathrm{A_{\phantom{\mbox{A}}}}} ^{2}}}$
        }
        \hbox{\vspace{1mm}}
        }
        &
        \vbox{
        \hbox{\vspace{3mm}}
        \hbox{
        $
        \tilde{k} _{\mu} = \cases{
                                \epsilon ^{\mu \alpha 0 1} k _{\alpha}
                                &
                                in the electric case
                                \cr
                                \cr
                                \epsilon ^{2 3 \mu \alpha} k _{\alpha}
                                &
                                in the magnetic case
                                 }
        $
        }
        \hbox{\vspace{1mm}}
        }
        \\ \hline
        \vbox{
        \hbox{\vspace{1mm}}
        \hbox{$k ^{2} _{\phantom{\mu}}$}
        \hbox{\vspace{1mm}}
        }
        &
        \vbox{
        \hbox{\vspace{3mm}}
        \hbox{
        $
        \tilde{k} ^{\perp} _{\mu}
        =
        \epsilon _{\mu \nu \alpha \beta}
        \bar{k} ^{\nu}
        \tilde{k} ^{\alpha}
        k ^{\beta}
        $
        }
        \hbox{\vspace{1mm}}
        }
        \\ \hline
        \vbox{
        \hbox{\vspace{1mm}}
        \hbox{$k ^{2} _{\phantom{\rho}}$}
        \hbox{\vspace{1mm}}
        }
        &
        \vbox{
        \hbox{\vspace{3mm}}
        \hbox{
        $
        E ^{\mu}
        =
        \epsilon ^{\mu \nu \rho \sigma}
        k _{\nu}
        \tilde{k} _{\rho}
        \tilde{k} ^{\perp} _{\sigma}
        $
        }
        \hbox{\vspace{1mm}}
        }
        \\
        \hline
\end{tabular}
\caption{\label{tab:eigen}Eigenvalues and eigenvectors of ${{\mathcal{D}} ^{-1}} ^{\mu \nu} (k)$.}%
}
There are, thus, two non-zero eigenvalues and one of them, $k ^{2}$,
has degeneracy two; moreover the ``gauge'' eigenvector is associated
to the zero eigenvalue. This last results changes if we study the
problem in the \textit{covariant} $\alpha$\textit{-gauge},
when the eigenvector $k ^{\mu}$ is then associated to a non-zero
(but still background independent) eigenvalue, $1/\alpha$ (the corresponding
inverse propagator will be called ${\mathcal{D} ^{-1}} ^{\mu \nu} ( k ; \alpha )$
as defined in what follows).

As promised above, we can now compare the obtained results with those derived in previous
works on the subject. Again, we observe that a key point in our derivation is the
(non-perturbative) path-integral procedure used to obtain equation (\ref{eq:NonPer}):
in this way all the effects due to the quantum fluctuations of the pseudoscalar field, at
all orders, are taken into account. In particular, if we concentrate on the purely
magnetic case, it is then clear that this constitutes a generalization of the results
obtained in \cite{bib:MaiPetZav}, where the secular equation is obtained considering
plane wave solutions to the \textit{classical} equations of motion associated to the
Maxwell $+$ Klein-Gordon action for the coupled electromagnetic and pseudoscalar fields.
An analogous result for the electric case in presence of massless pseudoscalars
can, for instance, be found in \cite{bib:ans}: here the dispersion relations are obtained
under physically very sensible restrictions but are, anyway, of perturbative character.
Also the more detailed analysis of \cite{bib:RafSto} uses a different kind of
approximation with respect to the one employed in our calculation since the starting
Lagrangian in equation (1) of \cite{bib:RafSto} is different from (\ref{eq:LquadraticElectric}),
our quadratic approximation to the full, non-perturbative, effective
result in equation (\ref{eq:NonPer}). We can thus trace back the differences between
our results for the eigenvalues of the propagator and the one already derived in the
literature on the subject, to the fact that we have taken into account the effects due
to the pseduscalar fields in a substantially non-perturbative way.

\section{\label{sec:proandvacene}Path integral quantization of small perturbations:
propagator and vacuum energy}

To compute the vacuum energy we proceed further in the covariantly quantized fashion
we started in the previous section. In a covariant $\alpha$-gauge the path integral
for the partition function can be rewritten, using (\ref{eq:opedef}) and going to
Euclidean space, as
\begin{eqnarray}
    Z
    & = &
    \left [ \det \left( \Box + m _{\mathrm{A}} ^{2} \right) \right] ^{-1/2}
    \int [ {\mathcal{D}} f ] \int [ {\mathcal{D}} A _{\mu} ]
        \delta \left[ \partial _{\mu} A ^{\mu} - f \right]
        e ^{- \frac{1}{2 \alpha} \int f ^{2} d ^{4} x}
        \times
        \det \left[ \partial ^{2} \right]
        \times
    \nonumber \\
    & & \quad
        \times
        \exp
        \left\{
            -
            \int \frac{d^4k}{(2\pi)^4}
                A _{\mu} (k)
                \left[
                    \left( k ^{2} g ^{\mu \nu} - k ^{\mu} k ^{\nu} \right)
                    -
                    \frac{
                        \kappa
                        \left(
                            \bar{k} ^{2} \bar{g} ^{\mu \nu}
                            -
                            \bar{k} ^{\mu} \bar{k} ^{\nu}
                        \right)}
                         {k ^{2} - m _{\mathrm{A}} ^{2}}
                \right]
                A _{\nu} (k)
        \right\}
    .
\end{eqnarray}
Here we have already performed the functional integration over $\phi$, we
remember that we are approximating the higher order Lagrangian up to the
second order in the fluctuations and understand the path-integrals
in the Euclidean sector. The functional integration over $f$ can now
be done and we get
\begin{equation}
    Z
    =
    \frac{
        \det \left[ \partial ^{2} \right]
        \int [ {\mathcal{D}} A _{\mu} ]
            \exp
            \left\{
                -
                \int \frac{d^4k}{(2\pi)^4}
                    A _{\mu} (k)
                    {{\mathcal{D}} ^{-1}} ^{\mu \nu} (k ; \alpha)
                    A _{\nu} (k)
            \right\}
         }
         {
        \left [ \det \left( \Box + m _{\mathrm{A}} ^{2} \right) \right] ^{1/2}
         }
    \label{eq:patintovevecpot}
    ,
\end{equation}
where
$$
    {{\mathcal{D}} ^{-1}} ^{\mu \nu} (k ; \alpha )
    =
    \left( k ^{2} g ^{\mu \nu} - k ^{\mu} k ^{\nu} \right)
    -
    \frac{\kappa}{k ^{2} - m _{\mathrm{A}} ^{2}}
    \left( \bar{k} ^{2} \bar{g} ^{\mu \nu}
    -
    \bar{k} ^{\mu} \bar{k} ^{\nu} \right)
    +
    \frac{1}{\alpha}
    k ^{\mu} k ^{\nu}
    =
    {{\mathcal{D}} ^{-1}} ^{\mu \nu} (k)
    +
    \frac{1}{\alpha} k ^{\mu} k ^{\nu}
    .
$$
In order to solve (\ref{eq:patintovevecpot}) we have to find the eigenvalues,
$\lambda _{k}$,
of ${{\mathcal{D}} ^{-1}} ^{\mu \nu} (k ; \alpha)$:
$Z$ equals then the product of these
eigenvalues. The requested eigenvalues are those related to the physical
polarizations found in section \ref{sec:sigmasgen}, i.e. $\tilde{k} _{\mu}$,
$\tilde{k} ^{\perp} _{\mu}$, $E _{\mu}$,
with eigenvalues
$k ^{2} - \kappa \bar{k} ^{2} / ( k ^{2} - m _{\mathrm{A}} ^{2} )$,
$k ^{2}$ and $k ^{2}$ respectively.
As already discussed at the end of the previous section,
the eigenvector $k ^{\mu}$ has now a non-zero eigenvalue
$1 / \alpha$.

The vacuum energy, or free energy, is then
$$
    W
    =
    \ln Z
    =
    \frac{1}{2}
    \sum _{j} \ln \lambda _{j}
    =
    \frac{V T}{2}
    \sum
    \int \frac{d^4k}{(2 \pi) ^{4}}
        \ln ( \lambda _{k} )
    ,
$$
so that
\begin{eqnarray}
    W
    & = &
    \frac{V T}{2}
    \int \frac{d^4k}{(2 \pi) ^{4}}
        \ln
        \left(
            k ^{2} - \frac{\kappa \bar{k} ^{2}}{k ^{2} - m _{\mathrm{A}} ^{2}}
        \right)
    +
    V T
    \int \frac{d^4k}{(2\pi)^4}
        \ln
        \left(
            k ^{2}
        \right)
    +
    \nonumber \\
    & & \qquad +
    \frac{V T}{2}
    \int \frac{d ^{4} k}{(2 \pi) ^{4}}
        \ln
        \left(
            k ^{2} - m _{\mathrm{A}} ^{2}
        \right)
    .
\label{eq:vaceneone}
\end{eqnarray}
The propagator, i.e. the inverse of ${{\mathcal{D}} ^{-1}} ^{\mu \nu} ( k ; \alpha )$,
can also be found exactly, as shown in appendix \ref{app:protecdet}, and results to be
\begin{eqnarray}
    & &
    {\mathcal{D}} ^{\mu \nu} ( k ; \alpha )
    =
    \frac{1}{k ^{2}}
    \left(
        g ^{\mu \nu} - \frac{k ^{\mu} k ^{\nu}}{k ^{2}}
    \right)
    +
    \frac{\kappa \bar{k} ^{2} \bar{g} ^{\mu \nu}}
         {
          k ^{2} \left( k ^{2} \left( k ^{2} - m _{\mathrm{A}} ^{2} \right)
          -
          \kappa \bar{k} ^{2} \right)
         }
    +
    \nonumber \\
    & & \qquad \qquad \qquad \qquad
    -
    \frac{\kappa \bar{k} ^{\mu} \bar{k} ^{\nu}}
         {
          k ^{2} \left( k ^{2} \left( k ^{2} - m _{\mathrm{A}} ^{2} \right)
          -
          \kappa \bar{k} ^{2} \right)
         }
    +
    \alpha
    \frac{k ^{\mu} k ^{\nu}}{( k ^{2} ) ^{2}}
    .
\end{eqnarray}

\section{\label{sec:vaceneeva}Evaluation of the vacuum energy and of its imaginary part}

To compute the vacuum energy, apart from more standard contributions,
we have to compute the following integral
\begin{equation}
    I _{0} ( k , \bar{k} ; \kappa , m _{\mathrm{A}})
    \equiv
    \int \frac{d^4k}{(2\pi)^4}
        \ln
        \left(
            k ^{2}
            -
            \frac{\kappa \bar{k} ^{2}}{k ^{2}
            -
            m _{\mathrm{A}} ^{2}}
        \right)
    .
\label{eq:nonstaint}
\end{equation}
It can be evaluated in closed form and we first rewrite it as
\begin{equation}
    I _{0} ( k , \bar{k} ; \kappa , m _{\mathrm{A}} )
    =
    I ( k , \bar{k} ; \kappa , m _{\mathrm{A}} )
    -
    \int \frac{d^4k}{(2\pi)^4}
        \ln
        \left(
            k ^{2} - m _{\mathrm{A}} ^{2}
        \right)
    ,
\label{eq:nonstaintsep}
\end{equation}
so that we can separate the second common contribution, from the first
one, i.e.
\begin{equation}
    I ( k , \bar{k} ; \kappa , m _{\mathrm{A}} )
    =
    \int \frac{d^4k}{(2\pi)^4}
        \ln
        \left[
            k ^{2} ( k ^{2} - m _{\mathrm{A}} ^{2} )
            -
            \kappa \bar{k} ^{2}
        \right]
    .
    \label{eq:nonstacon}
\end{equation}
Using (\ref{eq:nonstaintsep}) for the right hand side of (\ref{eq:nonstaint})
appearing in (\ref{eq:vaceneone}),
we get for the free energy
\begin{equation}
    W
    =
    \frac{V T}{2}
    I ( k , \bar{k} ; \kappa , m _{\mathrm{A}} )
    +
    V T
    \int \frac{d^4k}{(2\pi)^4}
        \ln
        \left(
            k ^{2}
        \right)
    \label{eq:vacenetwo}
    .
\end{equation}

The computation of $I ( k , \bar{k} ; \kappa , m _{\mathrm{A}} )$ is performed
in appendix \ref{app:vacenecomtecdet}. The infrared behavior
is then extracted in appendix \ref{app:inflimcomtecdet}. Ultraviolet
divergences  are regularized by a
cut-off $\Lambda$ and the leading contributions to the free energy
are computed in appendix \ref{app:ultlimcomtecdet}. Thus
the final result for the vacuum energy density
is obtained multiplying (\ref{eq:newparcon}) by $1 / ( 4 \pi ^{2} )$
(because of (\ref{eq:Jindef})) and substituting for
$I ( k , \bar{k} ; \kappa , m _{\mathrm{A}} )$ in (\ref{eq:vacenetwo}),
where the second contribution can also  be exactly evaluated.
The final result is
\begin{eqnarray}
    W
    =
    \frac{VT}{8 \pi ^{2}}
    \left [
        I _{\Lambda} ^{(4)} \Lambda ^{4}
        +
        I _{\ln \Lambda} ^{(4)} \Lambda ^{4} \ln \Lambda
        +
        I _{\Lambda} ^{(2)} \Lambda ^{2}
        +
        I _{\ln \Lambda} ^{(2)} \Lambda ^{2}
        +
        I ^{(0)}
        +
        I ^{(0)} _{\ln \Lambda}
    \right ]
    \label{eq:regeneden}
    ,
\end{eqnarray}
where the various contributions are defined at the end of appendix
\ref{app:vacenecomtecdet}. We will be especially interested in
$V T I ^{(0)} / (8 \pi ^{2})$, which, according to equation
(\ref{eq:fincorter}), is
\begin{eqnarray}
    I ^{(0)}
    & = &
    \frac{VT}{8 \pi ^{2}}
    \left [
        \frac{49 \kappa ^{2} + 132 \kappa m _{\mathrm{A}} ^{2} + 132 m _{\mathrm{A}} ^{4}}{576}
        +
        \frac{\kappa ^{2} + 3 \kappa m _{\mathrm{A}} ^{2} + 3 m _{\mathrm{A}} ^{4}}{48} \ln 2
        +
    \right.
    \nonumber \\
    & & \qquad
    \left .
        -
        \frac{( \kappa + m _{\mathrm{A}} ^{2} ) ^{3}}{24 \kappa}
        \ln
        \left(
            \kappa + m _{\mathrm{A}} ^{2}
        \right)
        +
        \frac{( m _{\mathrm{A}} ^{2} ) ^{3}}{24 \kappa}
        \ln
        \left(
            m _{\mathrm{A}} ^{2}
        \right)
        \right ]
    \label{eq:fincortertxt}
    .
\end{eqnarray}
From this expression we see
that the vacuum energy acquires an imaginary
part in the case $\kappa + m _{\mathrm{A}} ^{2} < 0$, i.e. when the
tachyonic modes are present. If we make use of the prescription
$\kappa + m _{\mathrm{A}} ^{2}$
$\to$
$\kappa + m _{\mathrm{A}} ^{2} + \imath \epsilon$,
the value of the imaginary part is
\begin{equation}
    - \pi
    \frac{V T}{ 8 \pi ^{2}}
    \frac{\left( \kappa + m _{\mathrm{A}} ^{2} \right) ^{3}}{24 \kappa}
    .
\label{eq:imaparvacene}
\end{equation}
Notice that, as opposed to the real part, the imaginary part of the vacuum
energy is cut-off independent. This is so because only the infrared
region of the integrand in (\ref{eq:nonstacon}) contributes
to the imaginary part.

We also observe that all contributions in (\ref{eq:regeneden}) are finite
in the limit of vanishing external field, i.e. when $\kappa \to 0$. This is
immediately evident for the terms $I _{\Lambda} ^{(4)}$, $I _{\ln \Lambda} ^{(4)}$,
$I _{\Lambda} ^{(2)}$, $I _{\ln \Lambda} ^{(2)}$ and $I _{\ln \Lambda} ^{(0)}$,
from their expressions at the end of appendix \ref{app:vacenecomtecdet}.
Concerning the contribution $I ^{(0)}$, the two divergent terms of opposite sign
in the second line of (\ref{eq:fincortertxt}) give a finite contribution in the limit,
as shown in (\ref{eq:Finliminf}).

The cut-off dependent parts are regularization dependent, as it is known
from general experience with the regularization of divergent integrals.
In particular subleading divergences (i.e. the $\Lambda ^{2}$ terms in
(\ref{eq:regeneden})) are highly dependent
upon the regularization scheme and may
vanish in certain of them.
In the present problem, due to the complexity of the integrand
in (\ref{eq:nonstacon}), other regularization schemes, like
$\zeta$-function regularization or dimensional regularization, are difficult
to implement.

\section{\label{sec:discon}Discussion and conclusions}

In this paper we have discussed effects of mass generation in an
external magnetic field and of tachyonic mass generation in an external electric field
in the case in which there is a pseudo scalar field with a pseudo scalar
coupling $g \phi F _{\mu \nu} \tilde{F} ^{\mu \nu}$.
The effects due to the presence of the psedo scalars are \textit{fully} considered:
indeed in the derivation of an effective theory for the fluctuations
of the electromagnetic field, the contributions from quantum fluctuations of
the pseudo scalars are taken into account using a \textit{non-perturbative} approach.

In the purely magnetic case the mass production for the electromagnetic fluctuations
can be interpreted as in the cited works already present in the literature: the
differences that we find in the eigenvalues of the propagator can be traced back
to the non-perturbative character of our approach,
as opposed to the perturbative analysis performed elsewhere.

A more careful discussion is instead required for the purely electric background,
in connection with what we have called \textit{tachyonic mass generation}.
Indeed the appearance of an imaginary part in the free energy suggests the presence
of an instability for homogeneous electric fields beyond some threshold due to
pseudo scalar coupling. This means that the vacuum state must be redefined,
to obtain a correct ground state for the theory. The analysis of what becomes the
true ground state, a genuinely non perturbative effect, is beyond the
scope of this manuscript. Nevertheless we would like, at least, to suggest a possible
and simple, although incomplete, answer to this question.
Indeed an analysis of Cornwall \cite{bib:Cor} indicates that in the case of a $3$-dimensional,
Euclidean, tachyonic mass term (which in our discussion is generated
by the time dependent scalar field of equation (\ref{eq:psescacoumas})
but can have also another origin) the final result is the formation of an
inhomogeneous state. Then, properly generalizing to our set-up the results described in
\cite{bib:GriMasMoh}, where inhomogeneous electromagnetic fields are shown to
decay in pseudoscalars, it is not unreasonable to understand under
what we have called \textit{tachyonic mass generation}
a real ``tachyonic instability of the vacuum''.

From the result of equation (\ref{eq:fincortertxt}) it is possible to see that
the free energy is well-defined in the limit of vanishing background fields
($\kappa \to 0$). Thus our effect is a genuinely non-perturbative one and does not
relate to a particular choice of the regularization scheme. Moreover
it is worth pointing out again that this ``tachyonic instability of the vacuum''
for fluctuations around a constant external electric field,
is characterized by a threshold effect, i.e. the tachyonic mass generation is switched on for electric
fields high enough, so that $\kappa + m _{\mathrm{A}} ^{2} < 0$.
In the case of the neutral pion, we can obtain the value of the effective
pion-photon coupling, defined by equation (\ref{eq:psescacoumas}),
from the observed value
of the neutral pion lifetime \cite{bib:pardatboo} and from the value of the decay rate
given the coupling (\ref{eq:psescacoumas}) \cite{bib:weiquathefie}.
This gives us the values
\begin{eqnarray}
    g
    & = &
    2.53 \cdot 10 ^{-5} \, \, \mathrm{MeV} ^{-1}
    ,
    \nonumber \\
    m _{\pi}
    & = &
    134.97 \, \, \mathrm{MeV}
    ,
    \nonumber \\
    \Longrightarrow E _{\mathrm{crit.}}
    & = &
    \left( 1 \, \, \mathrm{GeV} \right) ^{2}
    .
\label{eq:crielefieforpio}
\end{eqnarray}
This is a very high electric field,
not available in normal laboratory conditions.
Furthermore, if it were available,
it would reveal the composite structure of the
pion and the effective
$g \phi F _{\mu \nu} \tilde{F} ^{\mu \nu}$ coupling, used here, would
not be applicable any more.

A different question would be then the study of this effect
in the case of hypothetical axion particles. In this case, the threshold
for the tachyonic mass generation to be set up becomes lower as lower values for the mass
of the axion are considered.

Apart from the purely electric case, which is more subtle and, maybe, more \textit{exciting}
because of the \textit{exotic} tachyonic mass term, it could be interesting a more detailed
analysis of the magnetic case in connection with the set-up of PVLAS
a presently running experiment at the Legnaro I.N.F.N. laboratories, near Venice, Italy.

Finally a totally different role for these effects, could be in the context
of QCD. There, it is known that an external chromo-magnetic fields presents
tachyonic instability. If we were to add a particle
with coupling to
$\epsilon ^{\mu \nu \alpha \beta} F ^{a} _{\mu \nu} F ^{a} _{\alpha \beta}$
(this particle could represent a pseudo scalar bound state of quark and
anti-quark pairs), we know that the effect of the external
chromo-magnetic field together with the pseudo scalar coupling is of
generating mass.
The interplay of these two effects could then be an interesting
subject for further research.

\acknowledgments
We would like to thank Prof. John Cornwall for very interesting
comments and remarks about this work and Prof. Zvi Bern and
Prof. P. Gaete for conversations about the topics disussed here.
We also thank the referee for rising some points concerning the relation between the results
obtained in this paper and the ones already present in the literature on the subject and for
helping us, with his/her observations, in finding a mistake in the first version of the paper.
One of us, EIG, would like to thank University of Trieste and U.C.L.A. for hospitality
and I.N.F.N. for support.

\appendix

\section{\label{app:appdiaeig}Eigenvectors and
eigenvalues of ${{\mathcal{D}} ^{-1}} ^{\mu \nu} ( k )$}

We will now find the solutions to the eigenvector equation
\begin{equation}
    {{\mathcal{D}} ^{-1}} ^{\mu \nu} ( k )
    a _{\nu}
    =
    \lambda _{k}
    a ^{\mu}
    ,
\label{eq:eigequ}
\end{equation}
where ${{\mathcal{D}} ^{-1}} ^{\mu \nu} ( k )$ is defined in equation (\ref{eq:opedef}).

Firstly, there is a ``trivial'' gauge solution $a _{\mu} = H (k) k _{\mu}$,
since $k_\mu$ is orthogonal to both terms enclosed in round brackets
in the definition (\ref{eq:opedef}) of ${\mathcal{D}} ^{-1}$;
this can be seen from the relations
$$
    \left(
        k ^{2} g ^{\mu \nu} - k ^{\mu} k ^{\nu}
    \right)
    k _{\mu}
    =
    0
$$
and
$$
    \left(
        \bar{k} ^{2} \bar{g} ^{\mu \nu} - \bar{k} ^{\mu} \bar{k} ^{\nu}
    \right)
    k _{\mu}
    =
    0
$$
using the properties
$\bar{g} ^{\mu \nu} k _{\nu} = \bar{g} ^{\mu \nu} \bar{k} _{\nu}$
and
$\bar{k} ^{\nu} k _{\nu} = \bar{k} ^{\nu} \bar{k} _{\nu}$.

A second, non trivial polarization is
$\tilde{k} ^{\mu} = \epsilon ^{\mu \alpha 0 1} k _{\alpha}$
in the electric case
($\tilde{k} ^{\mu} = \epsilon ^{2 3 \mu \alpha} k _{\alpha}$
in the magnetic case),
since $\tilde{k} ^{\mu} k _{\mu} = 0 = \tilde{k} ^{\mu} \bar{k} _{\mu}$
thanks to equations (\ref{eq:norprodefuno}) and (\ref{eq:norprodefdue});
$a _{\mu} = \tilde{k} _{\mu}$ is associated to the eigenvalue
\begin{equation}
    k ^{2} - \frac{\kappa \bar{k} ^{2}}{k ^{2} - m _{\mathrm{A}} ^{2}}
    \label{eq:keytildisrel}
    .
\end{equation}

Then, remaining independent physically relevant polarizations must
be orthogonal to both $k _{\mu}$ and $\tilde{k} _{\mu}$, hence they can
be parametrized as
\begin{equation}
    a ^{\mu}
    =
    \epsilon ^{\mu \nu \alpha \beta}
    d _{\nu} \tilde{k} _{\alpha} k _{\beta}
    ;
\label{eq:othrelpol}
\end{equation}
since  $a ^{\mu}$ given by (\ref{eq:othrelpol}) is not affected by the
``gauge transformation''
$$
    d _{\nu}
    \longrightarrow
    d _{\nu}
    +
    \lambda _{1} \tilde{k} _{\nu}
    +
    \lambda _{2} k _{\nu}
    ,
$$
it then follows that only two components of $d _{\nu}$ are physically
relevant. One of these, which we will call $\tilde{k} ^{\perp} _{\mu}$,
is obtained when $d _{\nu} = \bar{k} _{\nu}$: it is thus orthogonal to
$k ^{\mu}$, $\bar{k} ^{\mu}$, $\tilde{k} _{\mu}$ and corresponds to the
eigenvalue $k ^{2}$. This is the same eigenvalue of the last eigenvector,
$E ^{\mu}$, which is given by
$$
    E ^{\mu}
    =
    \epsilon ^{\mu \nu \rho \sigma}
    k _{\nu}
    \tilde{k} _{\rho}
    \tilde{k} ^{\perp} _{\sigma}
    .
$$
This can be verified by inserting it into
(\ref{eq:fieequonesubFouspa}) and observing that
$$
    \bar{k} ^{2} \bar{g} _{\mu \nu}
    -
    \bar{k} _{\mu} \bar{k} _{\nu}
    \propto
    \tilde{k} _{\mu}
    \tilde{k} _{\nu}
    .
$$
The above equality holds because
$$
    \bar{k} ^{2} \bar{g} _{\mu \nu}
    -
    \bar{k} _{\mu} \bar{k} _{\nu}
$$
is a projector onto the space orthogonal to $\bar{k} _{\mu}$
\textit{in the $2$-dimensional subspace}, where the direction
orthogonal to $\bar{k} _{\mu}$ is nothing but $\tilde{k} _{\nu}$.

\section{\label{app:protecdet}Determination of the propagator}

To determine the propagator in the covariant $\alpha$-gauge,
${{\mathcal{D}} ^{-1}} ^{\mu \nu} ( k ; \alpha)$, with the four
dimensional quantities that are at our disposal we consider the
ansatz:
$$
    A
    \left(
        g ^{\mu \nu} - \frac{k ^{\mu} k ^{\nu}}{k ^{2}}
    \right)
    +
    B
    \bar{g} ^{\mu \nu}
    +
    C
    \bar{k} ^{\mu} \bar{k} ^{\nu}
    +
    D
    \frac{k ^{\mu} k ^{\nu}}{(k ^{2}) ^{2}}
    .
$$
To determine the coefficients $A$, $B$, $C$, $D$ we can now compute
${{\mathcal{D}} ^{-1}} ^{\mu \nu} ( k ; \alpha )
 {{\mathcal{D}} ^{-1}} _{\nu \alpha} ( k ; \alpha )$:
\begin{eqnarray}
    {\mathcal{D}} ^{-1 \, \mu \nu} {\mathcal{D}} _{\nu \alpha}
    & = &
    \left[
        \left(
            k ^{2} g ^{\mu \nu} - k ^{\mu} k ^{\nu}
        \right)
        -
        \frac{\kappa}{k ^{2} - m ^{2} _{\mathrm{A}}}
        \left(
            \bar{k} ^{2} \bar{g} ^{\mu \nu} - \bar{k} ^{\mu} \bar{k} ^{\nu}
        \right)
        +
        \frac{1}{\alpha}
        k ^{\mu} k ^{\nu}
    \right]
    \cdot
    \nonumber \\
    & & \qquad \cdot
    \left[
        A
        \left(
            g _{\nu \alpha} - \frac{k _{\nu} k _{\alpha}}{k ^{2}}
        \right)
        +
        B
        \bar{g} _{\nu \alpha}
        +
        C
        \bar{k} _{\nu} \bar{k} _{\alpha}
        +
        D
        \frac{k _{\nu} k _{\alpha}}{(k ^{2}) ^{2}}
    \right]
    \nonumber \\
    & = &
    A k ^{2} \delta ^{\mu} _{\alpha}
    -
    A k ^{\mu} k _{\alpha}
    -
    \frac{\kappa A}{k ^{2} - m ^{2} _{\mathrm{A}}}
    \left(
        \bar{k} ^{2} \bar{\delta} ^{\mu} _{\alpha}
        -
        \bar{k} ^{\mu} \bar{k} _{\alpha}
    \right)
    +
    \frac{A}{\alpha}
    k ^{\mu} k _{\alpha}
    +
    \nonumber \\
    & & \quad
    -
    A k ^{\mu} k _{\alpha}
    +
    A k ^{\mu} k _{\alpha}
    +
    \frac{\kappa A}{k ^{2} - m _{\mathrm{A}} ^{2}}
    \left(
        \frac{\bar{k} ^{2}}{k ^{2}} \bar{k} ^{\mu} k _{\alpha}
        -
        \frac{\bar{k} ^{2}}{k ^{2}} \bar{k} ^{\mu} k _{\alpha}
    \right)
    -
    \frac{A}{\alpha}
    k ^{\mu} k _{\alpha}
    +
    \nonumber \\
    & & \quad
    +
    B k ^{2} \bar{\delta} ^{\mu} _{\alpha}
    -
    B k ^{\mu} \bar{k} _{\alpha}
    -
    \frac{\kappa B}{k ^{2} - m ^{2} _{\mathrm{A}}}
    \left(
        \bar{k} ^{2} \bar{\delta} ^{\mu} _{\alpha}
        -
        \bar{k} ^{\mu} \bar{k} _{\alpha}
    \right)
    +
    \frac{B}{\alpha} k ^{\mu} \bar{k} _{\alpha}
    +
    \nonumber \\
    & & \quad
    +
    C k ^{2} \bar{k} ^{\mu} \bar{k} _{\alpha}
    -
    C \bar{k} ^{2} k ^{\mu} \bar{k} _{\alpha}
    -
    \frac{\kappa C}
         {k ^{2} - m ^{2} _{\mathrm{A}}}
    \left(
        \bar{k} ^{2} \bar{k} ^{\mu} \bar{k} _{\alpha}
        -
        \bar{k} ^{2} \bar{k} ^{\mu} \bar{k} _{\alpha}
    \right)
    +
    \frac{C}{\alpha}
    \bar{k} ^{2} k ^{\mu} \bar{k} _{\alpha}
    +
    \nonumber \\
    & & \quad
    +
    D \frac{k ^{\mu} k _{\alpha}}{k ^{2}}
    -
    D \frac{k ^{\mu} k _{\alpha}}{k ^{2}}
    -
    \frac{\kappa D}{k ^{2} - m _{\mathrm{A}} ^{2}}
    \left(
        \frac{\bar{k} ^{2}}{(k ^{2}) ^{2}} \bar{k} ^{\mu} k _{\alpha}
        -
        \frac{\bar{k} ^{2}}{(k ^{2}) ^{2}} \bar{k} ^{\mu} k _{\alpha}
    \right)
    +
    \frac{D}{\alpha}
    \frac{k ^{\mu} k _{\alpha}}{k ^{2}}
    .
    \nonumber
\end{eqnarray}
Now comparing terms with the same tensorial character
\begin{eqnarray}
    \delta ^{\mu} _{\alpha}
    & : &
    A k ^{2}
    =
    1
    \nonumber \\
    k ^{\mu} k _{\alpha}
    & : &
    \frac{D}{\alpha}
    \frac{1}{k ^{2}}
    =
    A
    \nonumber \\
    \bar{\delta} ^{\mu} _{\alpha}
    & : &
    B k ^{2}
    -
    \frac{\kappa (A + B) \bar{k} ^{2}}{k ^{2} - m _{\mathrm{A}} ^{2}} = 0
    \nonumber \\
    k ^{\mu} \bar{k} _{\alpha}
    & : &
    - B - C \bar{k} ^{2} = 0
    \nonumber \\
    \bar{k} ^{\mu} \bar{k} _{\alpha}
    & : &
    C k ^{2} - \frac{\kappa ( B + A )}{k ^{2} - m _{\mathrm{A}} ^{2}} = 0
    .
    \nonumber
\end{eqnarray}
An independent subset of (four of) these gives (consistently with the
remaining equation) the final result for the coefficients:
\begin{eqnarray}
    A & = & \frac{1}{k ^{2}}
    \nonumber \\
    B
    & = &
    \frac{\kappa \bar{k} ^{2} A}
         {
          k ^{2} \left( k ^{2} - m _{\mathrm{A}} ^{2} \right)
          -
          \kappa \bar{k} ^{2}
         }
    \nonumber \\
    \nonumber \\
    & = &
    \frac{\kappa \bar{k} ^{2}}
         {
          k ^{2} \left( k ^{2} \left( k ^{2} - m _{\mathrm{A}} ^{2} \right)
          -
          \kappa \bar{k} ^{2} \right)
         }
    \nonumber \\
    C
    & = &
    -
    \frac{\kappa}
         {
          k ^{2} \left( k ^{2} \left( k ^{2} - m _{\mathrm{A}} ^{2} \right)
          -
          \kappa \bar{k} ^{2} \right)
         }
    \nonumber \\
    D
    & = &
    \alpha
    ,
    \nonumber
\end{eqnarray}
so that we can finally write
\begin{eqnarray}
    & &
    {\mathcal{D}} ^{\mu \nu} ( k ; \alpha )
    =
    \frac{1}{k ^{2}}
    \left(
        g ^{\mu \nu} - \frac{k ^{\mu} k ^{\nu}}{k ^{2}}
    \right)
    +
    \frac{\kappa \bar{k} ^{2} \bar{g} ^{\mu \nu}}
         {
          k ^{2} \left( k ^{2} \left( k ^{2} - m _{\mathrm{A}} ^{2} \right)
          -
          \kappa \bar{k} ^{2} \right)
         }
    +
    \nonumber \\
    & & \qquad \qquad \qquad \qquad
    -
    \frac{\kappa \bar{k} ^{\mu} \bar{k} ^{\nu}}
         {
          k ^{2} \left( k ^{2} \left( k ^{2} - m _{\mathrm{A}} ^{2} \right)
          -
          \kappa \bar{k} ^{2} \right)
         }
    +
    \alpha
    \frac{k ^{\mu} k ^{\nu}}{(k ^{2}) ^{2}}
    .
    \nonumber
\end{eqnarray}

\section{\label{app:vacenecomtecdet}Explicit computation of the vacuum energy
integral}

We concentrate in this appendix on the computation of the contribution
$I ( k , \bar{k} ; \kappa , m)$ of equation (\ref{eq:nonstacon}). Of course since
the integral is divergent, it must be properly regularized and we choose to do that
by putting an infrared ($\epsilon$) and an ultraviolet ($\Lambda$) cutoff on the
modulus of the momentum $k$ and of its projection $\bar{k}$, exploiting some of
the arbitrariness in the choice of the regularization scheme. Thus integrals written
with implicit integration domain, like $\int d^{4} k (\dots)$, are to be understood as
performed in the domain of the variables $( k _{0}, k _{1}, k _{2}, k _{3} )$,
which is the inverse image of the domain
$\epsilon \leq r \leq \Lambda$, $0 \leq \vartheta < 2 \pi$,
$\epsilon \leq \rho \leq \Lambda$, $0 \leq \varpi < 2 \pi$
in the variables $(r , \rho , \vartheta , \varpi)$ under the following change of
variables in \textit{Euclidean} space:
$$
    \left(
        \matrix{k ^{4} \cr k ^{1} \cr k ^{2} \cr k ^{3}}
    \right)
    =
    T ( r , \vartheta , \rho , \varpi )
    =
    \left(
        \matrix{
            r \cos \vartheta
            \cr
            r \sin \vartheta
            \cr
            \rho \cos \varpi
            \cr
            \rho \sin \varpi
        }
    \right)
$$
with Jacobean
$$
    {\mathcal{J}} T
    =
    \left(
        \matrix{
            \cos \vartheta & - r \sin \vartheta & 0 & 0
            \cr
            \sin \vartheta & r \cos \vartheta & 0 & 0
            \cr
            0 & 0 & \cos \varpi & - \rho \sin \varpi
            \cr
            0 & 0 & \sin \varpi & \rho \cos \varpi
        }
    \right)
    ,
$$
whose determinant is
$$
    \det \left( {\mathcal{J}} T \right)
    =
    r \rho
$$
so that
$$
    d ^{4} k = r \rho d r d \rho d \vartheta d \varpi
    .
$$
We thus have
\begin{eqnarray}
    & &
    I ( k , \bar{k} ; \kappa , m _{\mathrm{A}} )
    =
    \nonumber \\
    & & \quad =
    \int \frac{d^4k}{(2\pi)^4}
        \ln
        \left[
            k ^{2} ( k ^{2} - m _{\mathrm{A}} ^{2}) - \kappa \bar{k} ^{2}
        \right]
    \nonumber \\
    & & \quad =
    \frac{1}{(2 \pi) ^{4}}
    \int _{0} ^{2 \pi} \! \! \! \! \! \! d \vartheta
    \int _{0} ^{2 \pi} \! \! \! \! \! \! d \varpi
    \int _{\epsilon} ^{\Lambda} \! \! \! \! \! \! d r
    \int _{\epsilon} ^{\Lambda} \! \!  \! \! \! \! d \rho
        \det \left( {\mathcal{J}} T \right)
        \times
        \ln
        \left[
            ( r ^{2} + \rho ^{2} )
            ( r ^{2} + \rho ^{2} - m _{\mathrm{A}} ^{2}) - \kappa \rho ^{2}
        \right]
    \nonumber \\
    & & \quad =
    \frac{1}{4 \pi ^{2}}
    \int _{\epsilon} ^{\Lambda} \! \! \! \! \! \! d r
    \int _{\epsilon} ^{\Lambda} \! \! \! \! \! \! d \rho
        \,
        r \rho
        \times
        \ln
        \left[
            r ^{4}
            +
            ( 2 \rho ^{2} - m ^{2} _{\mathrm{A}} ) r ^{2}
            +
            \rho ^{2} ( \rho ^{2} - m ^{2} _{\mathrm{A}} - \kappa )
        \right]
    \nonumber \\
    & & \quad \equiv
    \frac{1}{4 \pi ^{2}}
    \left .
        J ( r , \rho ; \kappa , m _{\mathrm{A}})
    \right | _{r , \rho = \epsilon} ^{r , \rho = \Lambda}
    \label{eq:Jindef}
    .
\end{eqnarray}
We have thus the problem of computing
$J ( r , \rho ; \kappa , m _{\mathrm{A}})$,
which can be calculated in closed form:
the final result can be expressed as
\begin{eqnarray}
    J ( r , \rho ; \kappa , m _{\mathrm{A}} )
    & = &
    \frac
    {
             \rho ^{2}
             \left[
                \kappa + 3 ( m _{\mathrm{A}} ^{2} - 6 r ^{2} -  \rho ^{2} )
             \right]
    }
    {24}
    +
    \nonumber \\
    \nonumber \\
    & & \quad
    +
    \frac{\BBB ^{3/2}}{48\kappa}
         \ln
         \left(
             \frac{\CCC - \sqrt{\BBB}}{\CCC + \sqrt{\BBB}}
         \right)
    -
    \frac{\AAA ^{3/2}}{48 \kappa}
    \ln
    \left(
        \frac{\kappa + \CCC - \sqrt{\AAA}}{\kappa + \CCC + \sqrt{\AAA}}
    \right)
    +
    \label{eq:IntRes} \\
    \nonumber \\
    & & \quad \quad \quad
    +
    \frac
    {
             \left[
                 \kappa ^{2}
                 +
                 3 \kappa (m _{\mathrm{A}} ^{2} - 2 r ^{2})
                 +
                 3 (m _{\mathrm{A}} ^{4} + 2 \DDD)
             \right]
    }
    {48}
    \ln
    \left(
        - \kappa \rho ^{2} + \DDD
    \right)
    \nonumber 
\end{eqnarray}
if we set
\begin{eqnarray}
    \AAA & = & A (r ; m _{\mathrm{A}} , \kappa)
      \equiv ( \kappa + m _{\mathrm{A}} ^{2} ) ^{2} - 4 \kappa r ^{2}
    \nonumber \\
    \BBB & = & B (\rho ; m _{\mathrm{A}} , \kappa)
      \equiv 4 \kappa \rho ^{2} + m _{\mathrm{A}} ^{4}
    \nonumber \\
    \CCC & = & C (r , \rho ; m _{\mathrm{A}})
      \equiv m _{\mathrm{A}} ^{2} - 2 ( r ^{2} + \rho ^{2} )
    \nonumber \\
    \DDD & = & D (r, \rho ; m _{\mathrm{A}})
      \equiv
      ( r ^{2} + \rho ^{2} ) ^{2}
      -
      m _{\mathrm{A}} ^{2} ( r ^{2} + \rho ^{2} )
      =
      \frac{\CCC ^{2} - \BBB}{4} + \kappa \rho ^{2}
    \nonumber
    .
\end{eqnarray}
The final stage consists in evaluating it in the infrared ($\epsilon \to 0$)
and ultraviolet limits. In this case this is equivalent to the computation of
$$
    \lim _{(r , \rho) \to (0 , 0)} J ( r , \rho ; \kappa , m _{\mathrm{A}} )
$$
and the extraction of the divergent contribution in
$$
    J ( r = \Lambda , \rho = \Lambda ; \kappa , m _{\mathrm{A}} )
$$
when $\Lambda \to \infty$. This is done in the next subsections.

\subsection{\label{app:inflimcomtecdet}Infrared limit}

In this subsection we consider the infrared limit and
in the following we will use the symbol ``$\cong$'' to imply that two
expression are equivalent in the infrared limit, i.e. they have the
same limit. Moreover we will get rid of the square roots by means of
the following results
\begin{eqnarray}
    \sqrt{
          1
          -
          \frac{4 \kappa r ^{2}}{( \kappa + m _{\mathrm{A}} ^{2} ) ^{2}}
         }
    & \cong &
    1 - \frac{2 \kappa r ^{2}}{| \kappa + m _{\mathrm{A}} ^{2} | ^{2}}
    \\
    \sqrt{1 + \frac{4 \kappa \rho ^{2}}{m _{\mathrm{A}} ^{4}}}
    & \cong &
    1 + \frac{2 \kappa \rho ^{2}}{m _{\mathrm{A}} ^{4}}
    .
\end{eqnarray}
We now turn to the contributions in the various lines of equation
(\ref{eq:IntRes}).
The first one gives no problem:
\begin{equation}
\frac
     {
      \rho ^{2}
      \left[
            \kappa + 3 ( m _{\mathrm{A}} ^{2} - 6 r ^{2} -  \rho ^{2} )
      \right]
     }
     {24}
     \cong
     0
     .
\label{eq:zerlimfir}
\end{equation}
The second one has a well behavior in the $\BBB$ term; more care has to
be paid in the logarithm:
\begin{eqnarray}
    \frac{\BBB ^{3/2}}{48\kappa}
    \ln
    \left(
         \frac{\CCC - \sqrt{\BBB}}{\CCC + \sqrt{\BBB}}
    \right)
    & \cong &
    \frac{m _{\mathrm{A}} ^{6}}{48 \kappa}
    \ln
    \left[
        \frac{
              m _{\mathrm{A}} ^{2}
              -
              2 ( r ^{2} + \rho ^{2} )
              -
              \sqrt{4 \kappa \rho ^{2}
              +
              m _{\mathrm{A}} ^{4}}
             }
             {
              m _{\mathrm{A}} ^{2}
              -
              2 ( r ^{2} + \rho ^{2} )
              +
              \sqrt{4 \kappa \rho ^{2}
              +
              m _{\mathrm{A}} ^{4}}
             }
    \right]
    \nonumber \\
    & \cong &
    \frac{m _{\mathrm{A}} ^{6}}{48 \kappa}
    \ln
    \left[
        \frac{
              m _{\mathrm{A}} ^{2}
              -
              2 ( r ^{2} + \rho ^{2} )
              -
              m _{\mathrm{A}} ^{2}
              \left(
                1 + \frac{2 \kappa \rho ^{2}}{m _{\mathrm{A}} ^{4}}
              \right)
             }
             {
              m _{\mathrm{A}} ^{2}
              -
              2 ( r ^{2} + \rho ^{2} )
              +
              m _{\mathrm{A}} ^{2}
              \left(
                1 + \frac{2 \kappa \rho ^{2}}{m _{\mathrm{A}} ^{4}}
              \right)
             }
    \right]
    \nonumber \\
    & \cong &
    \frac{m _{\mathrm{A}} ^{6}}{48 \kappa}
    \ln
    \left[
        - m _{\mathrm{A}} ^{2} ( r ^{2} + \rho ^{2} ) - \kappa \rho ^{2}
    \right]
    -
    \frac{m _{\mathrm{A}} ^{6}}{48 \kappa}
    \ln
    \left[
       {( m _{\mathrm{A}} ^{2} ) ^{2}}
    \right]
    .
\label{eq:zerlimsec}
\end{eqnarray}
The last term on the second line of (\ref{eq:IntRes}) again gives troubles only inside the
logarithmic term, which can be elaborated as follows
\begin{eqnarray}
    \ln
    \left(
        \frac{\kappa + \CCC - \sqrt{\AAA}}{\kappa + \CCC + \sqrt{\AAA}}
    \right)
    & = &
    \ln
    \left[
        \frac{
              \kappa + m _{\mathrm{A}} ^{2} - 2 ( r ^{2} + \rho ^{2} )
              -
              \sqrt{( \kappa + m _{\mathrm{A}} ^{2} ) ^{2} - 4 \kappa r ^{2}}
             }
             {
              \kappa + m _{\mathrm{A}} ^{2} - 2 ( r ^{2} + \rho ^{2} )
              +
              \sqrt{( \kappa + m _{\mathrm{A}} ^{2} ) ^{2} - 4 \kappa r ^{2}}
             }
    \right]
    \nonumber \\
    & \cong &
    \ln
    \left[
        \frac{
              \kappa + m _{\mathrm{A}} ^{2} - 2 ( r ^{2} + \rho ^{2} )
              -
              | \kappa + m _{\mathrm{A}} ^{2} |
              \left(
                1
                -
                \frac{2 \kappa r ^{2}}
                     {| \kappa + m _{\mathrm{A}} ^{2} | ^{2}}
              \right)
             }
             {
              \kappa + m _{\mathrm{A}} ^{2} - 2 ( r ^{2} + \rho ^{2} )
              +
              | \kappa + m _{\mathrm{A}} ^{2} |
              \left(
                1
                -
                \frac{2 \kappa r ^{2}}
                     {| \kappa + m _{\mathrm{A}} ^{2} | ^{2}}
              \right)
             }
    \right]
    \nonumber \\
    & \cong &
    \cases{
        \ln
        \left[
            \frac{
                  -
                  ( r ^{2} + \rho ^{2} )
                  +
                  \frac{\kappa r ^{2}}{\kappa + m _{\mathrm{A}} ^{2}}
                 }
                 {
                  ( \kappa + m _{\mathrm{A}} ^{2} )
                  -
                  ( r ^{2} + \rho ^{2} )
                  -
                  \frac{\kappa r ^{2}}{\kappa + m _{\mathrm{A}} ^{2}}
                 }
        \right]
        &
        if $\kappa + m _{\mathrm{A}} ^{2} > 0$
        \cr
        \ln
        \left[
            \left(
                \frac{
                      -
                      ( r ^{2} + \rho ^{2} )
                      +
                      \frac{\kappa r ^{2}}{\kappa + m _{\mathrm{A}} ^{2}}
                     }
                     {
                      ( \kappa + m _{\mathrm{A}} ^{2} )
                      -
                      ( r ^{2} + \rho ^{2} )
                      -
                      \frac{\kappa r ^{2}}{\kappa + m _{\mathrm{A}} ^{2}}
                     }
            \right) ^{-1}
        \right]
        &
        if $\kappa + m _{\mathrm{A}} ^{2} < 0$
    }
    \nonumber \\
    & \cong &
    \ln
    \left[
        \left(
            \frac{
                  -
                  ( r ^{2} + \rho ^{2} )
                  +
                  \frac{\kappa r ^{2}}{\kappa + m _{\mathrm{A}} ^{2}}
                 }
                 {
                  ( \kappa + m _{\mathrm{A}} ^{2} )
                  -
                  ( r ^{2} + \rho ^{2} )
                  -
                  \frac{\kappa r ^{2}}{\kappa + m _{\mathrm{A}} ^{2}}
                 }
        \right) ^{\mathrm{Sign} ( \kappa + m _{\mathrm{A}} ^{2} )}
    \right]
    \nonumber \\
    & \cong &
    \mathrm{Sign} ( \kappa + m _{\mathrm{A}} ^{2} )
    \!
    \left\{
        \ln \! \!
        \left[
            - m _{\mathrm{A}} ^{2} ( r ^{2} + \rho ^{2} ) - \kappa \rho ^{2}
        \right]
        \! - \!
        \ln \! \!
        \left[
            ( \kappa + m _{\mathrm{A}} ^{2} ) ^{2}
        \right]
    \right\}
    .
\label{eq:zerlimthib}
\end{eqnarray}
Since we also have
\begin{equation}
    \frac{\AAA ^{3/2}}{48 \kappa}
    \cong
    \frac{| \kappa + m ^{2} _{\mathrm{A}} | ^{3}}{48 \kappa}
    ,
\label{eq:zerlimthia}
\end{equation}
the two previous results, (\ref{eq:zerlimthia}) and (\ref{eq:zerlimthib}),
combine in a neat way: the sign in the first factor exactly combines with
the absolute value of the second factor
$$
    | \kappa + m _{\mathrm{A}} ^{2} | ^{3}
    \mathrm{Sign} ( \kappa + m _{\mathrm{A}} ^{2} )
    ,
$$
so that
\begin{eqnarray}
    & &
    \frac{\AAA ^{3/2}}{48 \kappa}
    \ln
    \left(
        \frac{\kappa + \CCC - \sqrt{\AAA}}{\kappa + \CCC + \sqrt{\AAA}}
    \right)
    \cong
    \nonumber \\
    & & \qquad \cong
    \frac{( \kappa + m _{\mathrm{A}} ^{2} ) ^{3}}{48 \kappa}
    \ln
    \left[
        - m _{\mathrm{A}} ^{2} ( r ^{2} + \rho ^{2} ) - \kappa \rho ^{2}
    \right]
    -
    \frac{( \kappa + m _{\mathrm{A}} ^{2} ) ^{3}}{48 \kappa}
    \ln
    \left[
        ( \kappa + m _{\mathrm{A}} ^{2} ) ^{2}
    \right]
    .
    \label{eq:zerlimthi}
\end{eqnarray}
For the last term in (\ref{eq:IntRes}) we do not have too much work.
The factor before the logarithm
has no problems and we can simply forget about the
$r$ and $\rho$ dependent parts.
Instead inside the logarithm we can neglect higher order terms in the limit
we are interested in, so that
\begin{eqnarray}
    & &
    \frac{
         \left[
              \kappa ^{2}
              +
              3 \kappa (m _{\mathrm{A}} ^{2} - 2 r ^{2})
              +
              3 (m _{\mathrm{A}} ^{4} + 2 \DDD)
         \right]
         }
         {48}
    \ln
    \left(
        - \kappa \rho ^{2} + \DDD
    \right)
    \cong
    \nonumber \\
    & & \qquad \cong
    \frac{
          \kappa ^{2}
          +
          3 \kappa m _{\mathrm{A}} ^{2}
          +
          3 m _{\mathrm{A}} ^{4}
         }
         {48}
    \ln
    \left[
        - m _{\mathrm{A}} ^{2} ( r ^{2} + \rho ^{2} ) - \kappa \rho ^{2}
    \right]
    .
    \label{eq:zerlimfou}
\end{eqnarray}
The desired result,
$J ( 0 , 0 ; \kappa , m _{\mathrm{A}})$ is then
(\ref{eq:zerlimfir})
$+$
(\ref{eq:zerlimsec})
$-$
(\ref{eq:zerlimthi})
$+$
(\ref{eq:zerlimfou}), i.e.
\begin{eqnarray}
    J ( 0 , 0 ; \kappa , m _{\mathrm{A}})
    & = &
    \frac{m _{\mathrm{A}} ^{6}}{48 \kappa}
    \left \{
        \ln
        \left[
            - m _{\mathrm{A}} ^{2} ( r ^{2} + \rho ^{2} ) - \kappa \rho ^{2}
        \right]
        -
        \ln
        \left[
            {( m _{\mathrm{A}} ^{2} ) ^{2}}
        \right]
    \right \}
    +
    \nonumber \\
    & & \quad
    -
    \frac{( \kappa + m _{\mathrm{A}} ^{2} ) ^{3}}{48 \kappa}
    \ln
    \left[
        - m _{\mathrm{A}} ^{2} ( r ^{2} + \rho ^{2} ) - \kappa \rho ^{2}
    \right]
    +
    \nonumber \\
    & & \quad
    +
    \frac{( \kappa + m _{\mathrm{A}} ^{2} ) ^{3}}{48 \kappa}
    \ln
    \left[
        ( \kappa + m _{\mathrm{A}} ^{2} ) ^{2}
    \right]
    +
    \nonumber \\
    & & \quad
    +
    \frac{
          \kappa ^{2}
          +
          3 \kappa m _{\mathrm{A}} ^{2}
          +
          3 m _{\mathrm{A}} ^{4}
         }
         {48}
    \ln
    \left[
        - m _{\mathrm{A}} ^{2} ( r ^{2} + \rho ^{2} ) - \kappa \rho ^{2}
    \right]
    \nonumber \\
    & = &
    \frac{( \kappa + m _{\mathrm{A}} ^{2} ) ^{3}}{24 \kappa}
    \ln
    \left(
        \kappa + m _{\mathrm{A}} ^{2}
    \right)
    -
    \frac{( m _{\mathrm{A}} ^{2} ) ^{3}}{24 \kappa}
    \ln
    \left(
        m _{\mathrm{A}} ^{2}
    \right)
    .
\label{eq:Intzerlim}
\end{eqnarray}
As also pointed out in the main text this contribution is \textit{finite}
in the case of vanishing external field ($\kappa \to 0$), since
\begin{equation}
    \lim _{\kappa \to 0}
        \left [
            \frac{( \kappa + m _{\mathrm{A}} ^{2} ) ^{3}}{24 \kappa}
            \ln
            \left(
                \kappa + m _{\mathrm{A}} ^{2}
            \right)
            -
            \frac{( m _{\mathrm{A}} ^{2} ) ^{3}}{24 \kappa}
            \ln
            \left(
                m _{\mathrm{A}} ^{2}
            \right)
        \right ]
    =
    \frac{m ^{4} _{\mathrm{A}}}{24}
    \left[
        \ln (m ^{6} _{\mathrm{A}}) + 1
    \right]
    .
\label{eq:Finliminf}
\end{equation}
\subsection{\label{app:ultlimcomtecdet}Ultraviolet limit}

To tackle the problem of the ultraviolet behavior of the energy density,
we analyze the limit in which $r \to \infty$, $\rho \to \infty$.
As already discussed we
will first set $r = \rho = \Lambda$ and then approximate the various
quantities as $\Lambda \to \infty$. For the relevant expressions, already
encountered above, we get\footnote{We now use the $\sim$ symbol to imply
\textit{the same behavior in the ultraviolet limit}.}
\begin{eqnarray}
    \AAA
    & = &
    -
    4 \kappa \Lambda ^{2}
    \left(
        1
        -
        \frac{( \kappa + m _{\mathrm{A}} ^{2} ) ^{2}}{4 \kappa \Lambda ^{2}}
    \right)
    ,
    \quad
    \AAA ^{2}
    \sim
    16 \kappa ^{2} \Lambda ^{4}
    \left(
        1
        -
        \frac{( \kappa + m _{\mathrm{A}} ^{2} ) ^{2}}{2 \kappa \Lambda ^{2}}
    \right)
    ,
    \quad
    \dots
    \label{eq:AAAapp} \\
    \BBB
    & = &
    4 \kappa \Lambda ^{2}
    \left(
        1
        +
        \frac{m _{\mathrm{A}} ^{4}}{4 \kappa \Lambda ^{2}}
    \right)
    ,
    \quad
    \BBB ^{2}
    \sim
    16 \kappa ^{2} \Lambda ^{4}
    \left(
        1
        +
        \frac{m _{\mathrm{A}} ^{4}}{2 \kappa \Lambda ^{2}}
    \right)
    ,
    \quad
    \dots
    \label{eq:BBBapp} \\
    \CCC
    & = &
    -
    4 \Lambda ^{2}
    \left(
        1
        -
        \frac{ m _{\mathrm{A}} ^{2}}{4 \Lambda ^{2}}
    \right)
    ,
    \quad
    \kappa + \CCC
    =
    -
    4 \Lambda ^{2}
    \left(
        1
        -
        \frac{ \kappa + m _{\mathrm{A}} ^{2}}{4 \Lambda ^{2}}
    \right)
    ,
    \nonumber \\
    & & \qquad \qquad
    ( \kappa + \CCC ) ^{-1}
    \sim
    -
    \frac{1}{4 \Lambda ^{2}}
    \left(
        1
        +
        \frac{ \kappa + m _{\mathrm{A}} ^{2}}{4 \Lambda ^{2}}
    \right)
    ,
    \quad
    \dots
    \label{eq:CCCapp} \\
    \DDD
    & = &
    4 \Lambda ^{4}
    \left(
        1
        -
        \frac{m _{\mathrm{A}} ^{2}}{2 \Lambda ^{2}}
    \right)
    ,
    \nonumber \\
    & & \qquad \qquad
    \log ( - \kappa \ \rho ^{2} + \DDD)
    \sim
    \log ( 4 \Lambda ^{4})
    -
    \frac{\kappa + 2 m _{\mathrm{A}} ^{2}}{4 \Lambda ^{2}}
    -
    \frac{( \kappa + 2 m _{\mathrm{A}} ^{2} ) ^{4}}{32 \Lambda ^{2}}
    .
    \label{eq:DDDapp}
\end{eqnarray}
Moreover we also have the well known expansions
\begin{eqnarray}
    {\mathrm{Arctanh}} \left( x \right)
    & = &
    x + \frac{x ^{3}}{3} + \frac{x ^{5}}{5} + {\mathcal{O}} (x ^{7})
    \label{eq:arctanhypapp} \\
    \arctan \left( x \right)
    & = &
    x - \frac{x ^{3}}{3} + \frac{x ^{5}}{5} + {\mathcal{O}} (x ^{7})
    ,
    \label{eq:arctanapp}
\end{eqnarray}
which we are going to use in the following. In particular we can consider in generality
the expansion of the following expression
\begin{eqnarray}
    -
    \frac{w ^{3/2}}{2}
    \ln
    \left(
        \frac{z - w ^{1/2}}{z + w ^{1/2}}
    \right)    
    & = &
    w ^{3/2}
    {\mathrm{Arctanh}} \left( \frac{w ^{1/2}}{z} \right)
    \nonumber \\
    & = &
    \cases{
        |w| ^{3/2}
        {\mathrm{Arctanh}}
        \left( \frac{|w| ^{1/2}}{z} \right)
        &
        if
        $w > 0$
        \cr
        \hfill \Rightarrow w ^{1/2} = |w| ^{1/2}
        &
        and$\quad w ^{3/2} = |w| ^{3/2}$
        \cr
        \cr
        -
        \imath |w| ^{3/2}
        {\mathrm{Arctanh}}
        \left( \frac{\imath |w| ^{1/2}}{z} \right)
        &
        if
        $w < 0$
        \cr
        \hfill \Rightarrow w ^{1/2} = \imath |w| ^{1/2}
        &
        and$\quad w ^{3/2} =  - \imath |w| ^{3/2}$
    }
    \nonumber \\
    & = &
    \cases{
        |w| ^{3/2}
        {\mathrm{Arctanh}}
        \left( \frac{|w| ^{1/2}}{z} \right)
        &
        if
        $w > 0$
        \cr
        \cr
        -
        \imath |w| ^{3/2}
        \imath \arctan
        \left( \frac{|w| ^{1/2}}{z} \right)
        &
        if
        $w < 0$
    }
    \nonumber \\
    & = &
    \cases{
        |w| ^{3/2}
        {\mathrm{Arctanh}}
        \left( \frac{|w| ^{1/2}}{z} \right)
        &
        if
        $w > 0$
        \cr
        \cr
        |w| ^{3/2}
        \arctan
        \left( \frac{|w| ^{1/2}}{z} \right)
        &
        if
        $w < 0$
    }
    \nonumber \\
    & = &
    |w| ^{3/2}
    \left [
        \left(
            \frac{|w| ^{1/2}}{z}
        \right)
        -
        {\mathrm{sign}} (w)
        \frac{1}{3}
        \left(
            \frac{|w| ^{1/2}}{z}
        \right) ^{3}
        +
    \right.
    \nonumber \\
    & & \qquad \qquad
    \left.
        +
        \frac{1}{5}
        \left(
            \frac{|w| ^{1/2}}{z}
        \right) ^{5}
        -
        {\mathrm{sign}} (w)
        \left(
            \frac{|w| ^{1/2}}{z}
        \right) ^{7}
        +
        \dots
    \right]
    \nonumber
    ,
\end{eqnarray}
so that
\begin{equation}
    -
    \frac{w ^{3/2}}{2}
    \ln
    \left(
        \frac{z - w ^{1/2}}{z + w ^{1/2}}
    \right)
    =
    w ^{2}
    \left[
        \sum _{n} ^{0 , \infty}
            \frac{w ^{n}}{(2 n + 1) z ^{2 n + 1}}
    \right]
    .
\label{eq:logappgen}
\end{equation}
From the above result, if we identify
\begin{eqnarray}
    w & \longleftrightarrow & \BBB ,
    \nonumber \\
    z & \longleftrightarrow & \CCC
    \nonumber
\end{eqnarray}
and we consider an overall $1/(24 \kappa)$ factor, using properly
(\ref{eq:BBBapp}) and the first equation of (\ref{eq:CCCapp}), we get
\begin{equation}
    \frac{\BBB ^{3/2}}{48 k}
    \ln
    \left(
        \frac{\CCC - \sqrt{\BBB}}{\CCC + \sqrt{\BBB}}
    \right)
    \sim
    \frac{\kappa \Lambda ^{2}}{6}
    +
    \frac{\kappa ^{2} + 3 \kappa m ^{2} + 6 m ^{4}}{72}
    \label{eq:secterapp}
\end{equation}
for the second term in (\ref{eq:IntRes}).
In the same way, starting again from result (\ref{eq:logappgen}),
together with the identifications
\begin{eqnarray}
    w & \longleftrightarrow & \AAA ,
    \nonumber \\
    z & \longleftrightarrow & \kappa + \CCC
    \nonumber
\end{eqnarray}
and taking into account an overall
$-1/(24 \kappa)$, (\ref{eq:AAAapp}) and (\ref{eq:CCCapp}),
we obtain for the third term in (\ref{eq:IntRes})
\begin{equation}
    -
    \frac{A ^{3/2}}{48 \kappa}
    \ln
    \left(
        \frac{\kappa + \CCC - \sqrt{\AAA}}{\kappa + \CCC + \sqrt{\AAA}}
    \right)    
    \sim
    -
    \frac{\kappa \Lambda ^{2}}{6}
    +
    \frac{4 \kappa ^{2} + 9 \kappa m ^{2} + 6 m ^{4}}{72}
    .
\label{eq:thiterapp}
\end{equation}
The last term in (\ref{eq:IntRes}) has also a logarithmic part
and, using (\ref{eq:DDDapp}), can be approximated as
\begin{eqnarray}
    & &
    \frac{1}{48}
    \left[
        \kappa ^{2} + 3 \kappa m _{\mathrm{A}} ^{2}
        - 6 \kappa r ^{2} + 3 m _{\mathrm{A}} ^{4}
        + 6 \DDD
    \right]
    \ln \left( - \kappa \rho ^{2} + \DDD \right)
    \sim
    \nonumber \\
    & & \qquad \qquad \sim
    \frac{\kappa ^{2} + 3 \kappa m _{\mathrm{A}} ^{2} + 3 m _{\mathrm{A}} ^{4}}{48} \ln \left( 4 \Lambda ^{4} \right)
    -
    \frac{\kappa + 2 m _{\mathrm{A}} ^{2}}{8} \Lambda ^{2} \ln \left( 4 \Lambda ^{4} \right)
    +
    \frac{\Lambda ^{4}}{2} \ln \left( 4 \Lambda ^{4} \right)
    +
    \nonumber \\
    & & \qquad \qquad \quad
    +
    \frac{( \kappa + 2 m _{\mathrm{A}} ^{2}) ^{2}}{64}
    -
    \frac{\kappa + 2 m _{\mathrm{A}} ^{2}}{8} \Lambda ^{2}
    .
    \label{eq:fouterapp}
\end{eqnarray}
We are now concerned with the easiest term in (\ref{eq:IntRes}), namely the first,
for which we get
\begin{equation}
    \frac{
          \rho ^{2}
          \left(
            \kappa
            +
            3
            \left( m _{\mathrm{A}} ^{2} - 6 r ^{2} - \rho ^{2} \right)
          \right)}
         {24}
    \sim
    \frac{\kappa + 3 m _{\mathrm{A}} ^{2}}{24} \Lambda ^{2}
    -
    \frac{7}{8} \Lambda ^{4}
    .
    \label{eq:firterapp}
\end{equation}
Summing up equations (\ref{eq:secterapp}),
(\ref{eq:thiterapp}), (\ref{eq:fouterapp}), (\ref{eq:firterapp}), we obtain
\begin{eqnarray}
    & &
    \Lambda ^{4}
    \left [
        2 \ln \Lambda
        +
        \ln 2
        -
        \frac{7}{8}
    \right ]
    +
    \Lambda ^{2}
    \left[
        -
        \frac{2 \kappa + 3 m _{\mathrm{A}} ^{2}}{24}
        -
        \frac{\kappa + 2 m _{\mathrm{A}} ^{2}}{4}
        \left(
            2 \ln \Lambda + \ln 2
        \right)
    \right]
    +
    \label{eq:infdiv}
    \\
    & & \qquad
    +
    \frac{5 \kappa ^{2} + 12 \kappa m _{\mathrm{A}} ^{2} + 12 m _{\mathrm{A}} ^{4}}{72}
    +
    \frac{(\kappa + 2 m _{\mathrm{A}} ^{2}) ^{2}}{64}
    +
    \frac{\kappa ^{2} + 3 \kappa m _{\mathrm{A}} ^{2} + 3 m _{\mathrm{A}} ^{4}}{48}
    \left(
         2 \ln \Lambda + \ln 2
    \right)
    .
    \nonumber
\end{eqnarray}

\subsection{Result for
$
    I ( r = \Lambda , \rho = \Lambda ; \kappa , m _{\mathrm{A}} )
    -
    \lim _{(r , \rho) \to (0 , 0)} I ( r , \rho ; \kappa , m _{\mathrm{A}} )
$
}

From the results (\ref{eq:Intzerlim}),
(\ref{eq:infdiv}) of the two previous subsections
we obtain what we are interested in,
\begin{eqnarray}
    & &
    J ( r = \Lambda , \rho = \Lambda ; \kappa , m _{\mathrm{A}} )
    -
    \lim _{(r , \rho) \to (0 , 0)} J ( r , \rho ; \kappa , m _{\mathrm{A}} )
    \sim
    \nonumber \\
    & & \qquad \sim
    \Lambda ^{4}
    \left [
        2 \ln \Lambda
        +
        \ln 2
        -
        \frac{7}{8}
    \right ]
    +
    \Lambda ^{2}
    \left[
        -
        \frac{2 \kappa + 3 m _{\mathrm{A}} ^{2}}{24}
        -
        \frac{\kappa + 2 m _{\mathrm{A}} ^{2}}{4}
        \left(
            2 \ln \Lambda + \ln 2
        \right)
    \right]
    +
    \nonumber \\
    & & \qquad \qquad
    +
    \frac{5 \kappa ^{2} + 12 \kappa m _{\mathrm{A}} ^{2} + 12 m _{\mathrm{A}} ^{4}}{72}
    +
    \frac{(\kappa + 2 m _{\mathrm{A}} ^{2}) ^{2}}{64}
    +
    \frac{\kappa ^{2} + 3 \kappa m _{\mathrm{A}} ^{2} + 3 m _{\mathrm{A}} ^{4}}{48}
    \left(
         2 \ln \Lambda + \ln 2
    \right)
    +
    \nonumber \\
    & & \qquad \qquad \qquad
    -
    \frac{( \kappa + m _{\mathrm{A}} ^{2} ) ^{3}}{24 \kappa}
    \ln
    \left(
        \kappa + m _{\mathrm{A}} ^{2}
    \right)
    +
    \frac{( m _{\mathrm{A}} ^{2} ) ^{3}}{24 \kappa}
    \ln
    \left(
        m _{\mathrm{A}} ^{2}
    \right)
    .
    \nonumber
    \nonumber \\
    & & \qquad \sim
    \left( I _{\Lambda} ^{(4)} - \ln 2 + \frac{3}{4} \right) \Lambda ^{4}
    +
    ( I _{\ln \Lambda} ^{(4)} - 2 ) \Lambda ^{4} \ln \Lambda
    +
    I _{\Lambda} ^{(2)} \Lambda ^{2}
    +
    I _{\ln \Lambda} ^{(2)} \Lambda ^{2}
    +
    I ^{(0)}
    +
    I ^{(0)} _{\ln \Lambda}
    \label{eq:newparcon}
\end{eqnarray}
where for convenience in the last line we have used the following
definitions:
\begin{eqnarray}
    I _{\Lambda} ^{(4)}
    & = &
    2 \ln 2 - \frac{13}{8}
    \nonumber \\
    I _{\ln \Lambda} ^{(4)}
    & = &
    4
    \nonumber \\    
    I _{\Lambda} ^{(2)}
    & = &
    -
    \frac{2 \kappa + 3 m _{\mathrm{A}} ^{2}}{24}
    -
    \frac{\kappa + 2 m _{\mathrm{A}} ^{2}}{4} \ln 2
    \nonumber \\
    I _{\ln \Lambda} ^{(2)}
    & = &
    -
    \frac{\kappa + 2 m _{\mathrm{A}} ^{2}}{2}
    \nonumber \\
    I ^{(0)}
    & = &
    \frac{49 \kappa ^{2} + 132 \kappa m _{\mathrm{A}} ^{2} + 132 m _{\mathrm{A}} ^{4}}{576}
    +
    \frac{\kappa ^{2} + 3 \kappa m _{\mathrm{A}} ^{2} + 3 m _{\mathrm{A}} ^{4}}{48} \ln 2
    +
    \nonumber \\
    & & \qquad
    -
    \frac{( \kappa + m _{\mathrm{A}} ^{2} ) ^{3}}{24 \kappa}
    \ln
    \left(
        \kappa + m _{\mathrm{A}} ^{2}
    \right)
    +
    \frac{( m _{\mathrm{A}} ^{2} ) ^{3}}{24 \kappa}
    \ln
    \left(
        m _{\mathrm{A}} ^{2}
    \right)
    \label{eq:fincorter} \\
    I ^{(0)} _{\ln \Lambda}
    & = &
    \frac{\kappa ^{2} + 3 \kappa m _{\mathrm{A}} ^{2} + 3 m _{\mathrm{A}} ^{4}}{24}
    \nonumber
    .
\end{eqnarray}

\end{document}